\documentclass[conference]{IEEEtran}
\usepackage{graphicx,epsfig,amsmath,amssymb,bm}
\usepackage{amssymb,balance}
\usepackage{amsmath}
\usepackage{amsthm} 
\usepackage{amssymb}
\usepackage{algorithmicx}
\usepackage{algpseudocode}
\usepackage{cite}
\usepackage{url}
\newfont{\bbb}{msbm10 scaled 500}

\newfont{\bb}{msbm10 scaled 1100}

\usepackage{bbm}
\usepackage{subcaption}
\usepackage{tabularx}
\usepackage{mathtools}
\usepackage{empheq}
\usepackage{multirow}
\usepackage{verbatim}
\usepackage[ruled,linesnumbered]{algorithm2e}
\usepackage{color}
\usepackage{tikz}
\usepackage[acronym]{glossaries}
\usepackage[capitalise,noabbrev]{cleveref}
\usepackage{microtype}
\include{defn}
\usepackage{enumitem}

\newcommand{\beqa}{\begin{eqnarray}}
\newcommand{\eeqa}{\end{eqnarray}}

\ifCLASSINFOpdf

\else

\fi

\begin{document}

\title{Analysis of Cascading Failures Due to Dynamic Load-Altering Attacks \vspace{-0.3 cm}}
\IEEEoverridecommandlockouts % to enable \thanks in authorlist
\author{
\IEEEauthorblockN{Maldon Patrice Goodridge\IEEEauthorrefmark{1}, Alessandro Zocca\IEEEauthorrefmark{2}, Subhash Lakshminarayana\IEEEauthorrefmark{3}  } 
\IEEEauthorblockA{\IEEEauthorrefmark{1}Global Development Initiatives  \\
\IEEEauthorrefmark{2}Department of Mathematics, Vrije Universiteit Amsterdam, NL\\
\IEEEauthorrefmark{3}School of Engineering, University of Warwick, UK \\
Emails: \IEEEauthorrefmark{1}gdi.uklimited@gmail.com, 
\IEEEauthorrefmark{2}a.zocca@vu.nl, \IEEEauthorrefmark{3}subhash.lakshminarayana@warwick.ac.uk
}
\vspace{-0.9 cm}
}

% \thanks{
% This work was supported in part by a startup grant at the University of Warwick and in part by the U.S. National Science Foundation under Grants  DMS-1736417 and ECCS-1824710.}

% \title{A Rare-Event Sampling Approach for the Analysis of Load-Altering Attacks Against Power Grids}
% \author{Maldon Patrice Goodridge and Subhash~Lakshminarayana~\IEEEmembership{Senior Member, IEEE},  \vspace{-0.2in}\thanks{S. Lakshminarayana is with the School of Engineering, University of Warwick, Coventry, UK, CV47AL (email: subhash.lakshminarayana@warwick.ac.uk). 
% }
% }

% make the title area
\maketitle

\begin{abstract}
Large-scale load-altering attacks (LAAs) are known to severely disrupt power grid operations by manipulating several internet-of-things (IoT)-enabled load devices. In this work, we analyze power grid cascading failures induced by such attacks. The inherent security features in power grids such as the $N-1$ design philosophy dictate LAAs that can trigger cascading failures are \emph{rare} events.
We overcome the challenge of efficiently sampling critical LAAs scenarios for a wide range of attack parameters by using the so-called ``skipping sampler'' algorithm. We conduct extensive simulations using a three-area IEEE-39 bus system and provide several novel insights into the composition of cascades due to LAAs. Our results highlight the particular risks to modern power systems posed by strategically designed coordinated  LAAs that exploit their structural and real-time operating characteristics. 
\end{abstract}

\IEEEpeerreviewmaketitle

\section{Introduction}
% indirect attacks that target a large number of demand-side appliances in a Botnet-type attack have been studied only recently \cite{Dabrowski2017, Soltan2018}. Unlike the SCADA assets, these devices cannot be monitored continuously due to their large numbers. 

% Use for journal version
%The BlackEnergy attack against Ukraine’s power grid in December 2015 caused power outages for more than 225,000 customers for several hours  \cite{Ukraine2016:Analysis}.

%{\bf Maldon: can you include economic cost of this attack, i think this will strengthen the argument}

% The balance between generation and demand is critical for power grid operations. 
Load-altering attacks (LAAs) can cause sudden changes in power grid demand by compromising tens of thousands of internet-of-things (IoT) enabled high-wattage electrical appliances (smart heat pumps, electric vehicle charging stations), and can disrupt the power equilibrium and threaten system safety \cite{Dabrowski2017, Soltan2018}. The growing penetration of renewable energy resources resulting in low-inertia conditions can exacerbate the consequences of such attacks \cite{LakshCOVID2022}. 

% LAAs have attracted significant attention in the literature. 
LAAs can be broadly divided into two categories \cite{LakshIoT2021} -- (i) static LAAs (S-LAAs) and (ii) dynamic LAAs (D-LAAs). S-LAAs refer to a sudden, one-time change in loads \cite{Soltan2018}, whereas, D-LAAs refer to a series of load changes over a period of time \cite{AminiLAA2018}. S-LAAs can result in network frequency and/or line flows exceeding safety limits, leading to component disconnections \cite{Soltan2018, HuangUSENIX2019}. If an attacker changes system loads proportionally to the frequency deviations, they can potentially destabilize the frequency control loop, leading to cascading failures. An analytical approach to studying the effects of static/dynamic LAAs was proposed in \cite{LakshIoT2021} using the theory of second-order dynamical systems and identifying the nodes from which an attacker can launch the most impactful attacks. 

%They can also result in transmission line power flows violating the safety limits, leading to cascading failures.

%% Text on defense, use if necessary %%
%% Subsequent work has also investigated methods to defend against LAAs. In particular, offline methods such as the optimal deployment of security features in load devices (e.g., encryption-enabled smart appliances) and determining generator set-points to prevent the destabilizing effects of LAA were investigated in \cite{AminiLAA2018} and \cite{**} \ale{missing ref!} respectively. However, given the lack of unified security standards, offline defense may not be practical. Recent work also proposed online detection and localization of LAAs by monitoring voltage and frequency data from phasor measurement units following LAAs in \cite{AminiIdentification2019} and \cite{lakshminarayana2021datadriven}.
%%%

% Subsequent work has also focused on detecting LAAs using data-driven approaches based on the data gathered from  \cite{AminiIdentification2019, lakshminarayana2021datadriven}.

% Most of the existing work on this topic takes a simulation-based appra

% Existing work on this topic has focussed on attack impact analysis and defending against such attacks. 

% Defending against LAAs is challenging, given that these load devices are typically installed and operated at consumer sites.  

Despite the growing literature on LAAs, existing research lacks a framework to understand the extent of consequences, specifically, in terms of the cascading disconnections such large-scale attacks can cause. Reference \cite{Soltan2018} was the first to investigate this direction. However, their analysis did not consider power grid protection features such as $N-1$ security and load shedding. As a result, they significantly overestimated the extent of cascades. Reference \cite{HuangUSENIX2019} analyzed LAA-induced cascading failures considering the aforementioned security features. However, the results presented in both \cite{Soltan2018} and \cite{HuangUSENIX2019} correspond to only a few specific LAA scenarios (i.e., specific load perturbations injected at a few of the victim nodes). They do not provide a thorough understanding of the distribution of cascades due to all possible spatial LAA scenarios over the victim nodes.

% Additionally, while many analyses of cascading failures in the literature focus on the disconnection of lines and loads in a network \cite{yao2016cascades,stumercascades2021}, we instead investigate the loss of both active generation and loads following an LAA. This accounts for an attacker's intention to impair any network operation to inflict tactical, economic, or social costs.

However, identifying LAAs that lead to cascades is challenging, since $N-1$ design philosophies ensure that the power grid is resilient to the destabilizing effects of such load changes \cite{HuangUSENIX2019}. Thus, LAA instances that lead to cascading failures are in fact \emph{rare events} and sampling them efficiently for a wide range of attack parameters becomes highly nontrivial. Despite their low likelihood, rare events are key to understanding power systems reliability, see e.g.~\cite{Nesti2018}.

In the literature, simulating cascading failures involves using complex models which require significant computational resources (e.g. \cite{song2016cascades,stumercascades2021}), rendering such models unsuitable for rare event identification. Instead, we pair a fast-evaluating dynamic model with a sampling methodology to identify network and LAA parameters associated with cascading failures, building on the methodology presented in \cite{goodridge2020}. Specifically, we make use of the \textit{skipping sampler}, a Markov Chain Monte Carlo (MCMC) sampling algorithm specifically designed to efficiently sample low-likelihood events. The core idea behind this method is to traverse ``uninteresting'' regions of the parameter space (i.e. ``skip'') in a systematic way to more efficiently sample rare but critical D-LAA characteristics.%hence managing to discover and sample more efficiently all the successful D-LAA modes. 
The skipping sampler has already been successfully used in the context of power systems security in~\cite{GoodridgePMAPS2022} to identify S-LAAs that lead to the activation of emergency responses and in~\cite{moriarty2018frequency} to understand the risks of correlated frequency violations.

%which combined a dynamic power system model with a Monte Carlo sampler to investigate cascade sizes in simplified, stylized power systems.

We significantly generalize the framework proposed in~\cite{GoodridgePMAPS2022} and extend it to analyse LAA-induced cascades. Specifically, as opposed to the exclusive focus on S-LAAs considered in \cite{GoodridgePMAPS2022}, we instead present a framework that captures both static and dynamic LAAs in a unified manner. We model attacks that inject a sequence of periodic load alterations proportional to the frequency deviation (see \cite{AminiLAA2018}). By varying the interval between attacks, we can move from a static attack model with long intervals between attacks to a dynamic one with frequent attacks. Lastly, we consider the sequence of power grid emergency responses, and how they cascade over the simulation period and investigate their distributions as a function of various attack parameters.  
We perform extensive simulations using a three-area IEEE-39 bus system and provide the following novel insights into the composition of cascades induced by LAAs. Specific contributions include:
\begin{itemize}[leftmargin=.5cm,labelsep=0.2cm]
    \item Analysis of power grid cascades as a function of the attack parameters, namely (i) the amount of vulnerable load accessible by the attacker, and (ii) the interval between successive load attacks (for DLAA).
    \item Analysis of cascades due to LAAs under different loading conditions and inter-area power balance.
    \item Identification of dominant failure modes under different attack and power grid operational regimes. 
\end{itemize}
Our results show that there are two attack regimes in which the grid is particularly vulnerable, (i) low-magnitude attacks with a short interval between the load changes (D-LAAs), and (ii) very large-magnitude attacks with a long interval between the load changes (S-LAAs).  Furthermore, the network is highly vulnerable to D-LAAs in peak demand periods and areas with large power imbalances are particularly vulnerable to cascading failures due to D-LAAs.

The rest of the paper is organized as follows. \cref{sec:Prelim} introduces the system model; \cref{sc:statmodel} presents the rare-event sampling algorithm; \cref{sec:Sims} describes the simulation results and \cref{sec:Conc} concludes.

 %described in \cite{goodridge2021} and \cite{GoodridgePMAPS2022}
 % It also models frequency protection systems, which disconnect generators, loads, and inter-area transmission lines when frequency-related profiles exceed pre-defined operational thresholds. We defer the interested reader to~\cite{goodridge2021} for more detailed specifications of the model and the parameters.
 
\section{System model}\label{sec:Prelim}
Using a graph theoretic formulation, the power system can be described as $\mathcal{S} = \{ \mathcal{N}, \mathcal{W} \},$ with $\mathcal{N}$ is the set of buses and $\mathcal{W}$ is the set of transmission lines. We decompose $\mathcal{N} = \mathcal{G} \cup \mathcal{L},$ where $\mathcal{G}$ is the set of generator buses and $\mathcal{L}$ is the set of load buses. 
The evolution of power grid dynamics is modelled using a third-order model, which models frequency and voltage transients following a power injection in the network, as well as generator governor action and automatic voltage regulation.
Unde this model, for each generator $i \in \{1, \dots N\}$, dynamics in voltage phase angle $\delta_i$, voltage magnitude $E_i$ and governor action $\rho_i$ are given respectively by:

{\small
\begin{subequations}
\label{eq:network_model}
\begin{empheq}[left = \empheqlbrace\,]{flalign}
%	\begin{flalign}
        & M(\psi) \ddot{\delta}_{i}+D \dot{\delta}_{i}=\psi_i \chi^G_i- \chi^L_i(\mathcal{R}_i)\nonumber \\ &\hspace{3cm} -E_{i}\sum_{j=1}^{N+L}B_{ij}(\Omega_{ij})E_{j}\sin(\delta_{ij} )\label{eq:freq21} \\
        & S_{i}\dot{E}_{i}=\psi_i (E_{f,i}-\text{v}_i) -E_{i}+ X_{i}\sum_{j=1}^{N+L}B_{ij}(\Omega_{ij})E_{j}\cos(\delta_{ij} )\label{eq:volt21}\\
        & \dot{\rho}_{i}= -A_{i}\dot{\delta}_{i}(1-1_{\mathcal{\mathcal{\mathcal{W}}}}[\dot{\delta}_{i}]).\label{eq:gov21}
    %\end{flalign}
    \end{empheq}
\end{subequations}}
% %=
\noindent Similarly, the dynamics for $\delta_i$ and $E_i$ at each load bus $i \in \{N+1, \dots, N+L\}$ are given by the following system of equations: 

{\small
\begin{subequations}
\label{eq:network_model_load}
% 	\begin{align}%{align}
\begin{empheq}[left = \empheqlbrace\,]{flalign}
        & M(\psi) \ddot{\delta}_{i}+D \dot{\delta}_{i}= -\chi^L_i(\mathcal{R}_i)- E_{i}\sum_{j=1}^{N+L}B_{ij}(\Omega_{ij})E_{j}\sin(\delta_{ij})\label{eq:freq_load} \\
        & S_{i}\dot{E}_{i}=\psi_i E_{f,i}-E_{i}+X_{i}\sum_{j=1}^{N+L}B_{ij}(\Omega_{ij})E_{j}\cos(\delta_{ij} )\label{eq:volt_load}
% 	\end{align}
\end{empheq}
\end{subequations}}
In equations~\eqref{eq:network_model} and~\eqref{eq:network_model_load}, $\psi_i$, $\Omega_{ij}$ and $\mathcal{R}_i$ are indicator variables for frequency protection models and reflect the disconnection of network components- i.e.- generators, lines and loads respectively (see \cref{sec:emer_res}). 
% The network's angular momentum is calculated as the sum of each generator's inertia constant $H_i$, that is, $M(\psi)= \sum_{j=1}^{N} \psi_j H_j$, accounting for the cases where generators may be disconnected ($\phi_j =0$).
 \begin{table}[!h]
    \centering
    \caption{Variables used in~\eqref{eq:network_model} and~\eqref{eq:network_model_load}. }
    \begin{tabular}{|c|l|c|}
    \hline
    \textbf{Symbol}&\textbf{Meaning}&\textbf{Units}\\\hline
     $A_i$&Governor's droop response &MW/rad\\                 $B_{ij}(\Omega_{ij})$&Susceptance matrix&p.u.\\   $\chi_i^G$&Net generation at node $i$&p.u.\\
     $\chi_i^L(\mathcal{R})$&Net loads at node $i$&p.u.\\
     $D$&System damping& \%\\
     $\delta_i$ & Phase angle & p.u\\
     $\delta_{ij}$&$\delta_i - \delta_j$&p.u.\\
     $\dot{\delta}_i$&Frequency & p.u\\
     $\ddot{\delta}_i$&Rate of change of frequency (RoCoF)&p.u.\\
     $E_i$ & Voltage&p.u.\\ 
     $E_{f,i}$&Machine $i$ rotor field voltage&p.u.\\         $M(\psi)$&System angular momentum & Ws$^2$\\
     $\Omega_{ij}$&Line disconnection indicator&-\\
     $\psi_i$ & Generator shed indicator&-\\
     $\mathcal{R}_i$&UFLS counter&-\\
     $S_i$&Machine $i$ transient time constant&s\\
     $X_i$&Machine $i$ equivalent reactance&ohms\\
     $\mathcal{W}$&Governor's deadband frequency range&Hz\\
     \hline
\end{tabular}
\label{tb:model_variables}
\end{table}
 Net power injection at nodes $i\in{1,\dots,N}$ is given by $\chi^G_i = \min\{P_i^{\text{max}},P^G_i + \rho_i\}$, where $P_i^{\text{max}}$ is the stated maximum power output of generator $i$, $P_i^G$ is the power of the generator in equilibrium and $\rho_i$ is the power contributed by a governor unit \eqref{eq:gov21} \cite{GoodridgePMAPS2022}.  The variable $v_i$ represents the actions of automatic voltage regulation (see online Appendix). The net load at node $i$, $\chi^L_i$, includes equilibrium loads, the dynamic LAA and a load disconnection scheme, and is discussed in \cref{sec: laa_model,sec:emer_res}. The remaining parameters are given in \cref{tb:model_variables}.
%  \vspace{-0.15cm}
\subsection{Dynamic load-altering attack model}
\label{sec: laa_model}
%  \vspace{-0.1cm}
% Following \cite{goodridge2022}, nodal equilibrium loads $P^L_i$ are decomposed based on their vulnerability to LAAs, such that $P^L_i = (1-\nu) P^L_i + \nu P^L_i$, where the network parameter $\nu \in [0,1]$ represents the proportion of nodal loads susceptible to cyber-manipulation.
% A dynamic load-altering attack can be modelled as a time-dependent sequence of changes to the vulnerable loads $\nu P^L_i$ at each node $\lambda_i = \{\lambda_{i,0},\lambda_{i,1},\dots,\lambda_{i,h}\} $, at discrete times $t_k=k\mathcal{I}$ for **$k =0, 1, 2,\dots, h$** with fixed interval $\mathcal{I}$ seconds over the course of the simulation. Each load change is applied as an instantaneous impulse and kept constant until the next update in the sequence. The initial load change, $\lambda_0$ at $t=0$, is selected randomly by the sampling procedure (see section XX). Subsequent load changes $\lambda_k$ for $k\ge 1$ are deterministic and calculated as a change to vulnerable loads as network dynamics evolve using a frequency-dependent `reverse governor' model. This describes an attacker intent  with access to quasi-real-time or simulated frequency data,  intent on using the sequence of attacks to exacerbate frequency deviations.
\vspace{-0.1 cm}
We denote the maximum load at each node $i \in \mathcal{L}$ by $P^L_{i,\max}.$ At the start of the simulation (effectively modelling different times of the day, see Section~IV), let us denote the load at node $i \in \mathcal{L}$ by $P^L_i \in [0,P^L_{i,\max}].$ A D-LAA can be modelled as a time-dependent sequence of load changes at the vulnerable load nodes. More specifically, we denote by $\lambda_{i}(t_k) \in \mathbb{R}$ the magnitude in MW of the LAA at node $i$ occurring at time epoch $t_k$ for $k \in \mathbb{N}$. We assume that the time epochs are separated by a pre-determined, constant time interval $\mathcal{I}$. Each load change is applied as an instantaneous impulse and kept constant until the next update in the sequence. The initial load change, $\lambda_i(t_0)$, is selected randomly by the sampling procedure (see~\cref{sc:statmodel}). Subsequent load changes $\lambda_i(t_k)$ for $k\ge 1$ are deterministic, calculated using a frequency-dependent `reverse governor' model. This describes an attacker with access to network data, intent on using the sequence of attacks to exacerbate frequency deviations (the premise of D-LAAs \cite{AminiLAA2018}). In mathematical terms, for every vulnerable load node $i$ and for every index $k \geq 1$
\begin{equation} \label{eqn:DLAA_main}
    \lambda_{i}(t_k) = 
    \begin{cases}
            C\dot{\delta}_i(t_k) &  \chi^{L-}_i-C\dot{\delta}_i(t_k) \in [0,P^L_{i,\text{max}}],\\
            P^L_{\text{max}} - \chi^{L-}_i & \chi^{L-}_i-C\dot{\delta}_i(t_k) >P^L_{i,\text{max}},\\
            -\chi^{L-}_i& \chi^{L-}_i-C\dot{\delta}_i(t_k) <0,
    \end{cases}
\end{equation}
\noindent where $\lambda_i(t_k) $ is the $k^{th}$ D-LAA component at time $t_k = k\mathcal{I}$ for $k\in \mathbb{N}$, $\dot{\delta}_i(t_k)$ is the frequency at node $i$ at time $t=t_k$, $C\in \mathbb{R}^+$ is a network variable which relates the change in load to the frequency deviation at $t = t_k$. The expression $\chi^{L-}_i = P^L_i +\sum_{j=0}^{k-1}\lambda_i(t_j)$ refers to the total net loads at node $i$ just before the application of the $k^{\mathrm{th}}$ load change, inclusive of all previous load changes and load shedding. 

In essence, \eqref{eqn:DLAA_main} ensures that the load at node $i \in \mathcal{L}$ following the D-LAA remains within the set $[0,P^L_{i,\text{max}}].$ The framework presented above models S- and D-LAAs in a unified manner, specifically by varying $t_k.$ Note that $t_k \to 0$ represents a continuous load attack (i.e., the D-LAA framework presented in \cite{AminiLAA2018}), while $t_k \to \infty$  models an S-LAA (i.e., only a single load attack over the entire simulation interval). The D-LAA at each node can be aggregated in a straightforward manner, giving rise to two key metrics:
\begin{enumerate}[leftmargin=.5cm,labelsep=0.2cm]
\item The \textit{cumulative D-LAA} (in MW),  $\Sigma_i(\lambda) := \sum_k |\lambda_{i}(t_k)| \in \mathbb{R}^L$, which measures the size of the attack at node $i$ as the sum of the magnitudes of D-LAAs at that node over the duration of the simulation. 
\item The \textit{average network load change} (in MW); $\mu_i(\lambda,h) := \sum_{t=1}^h (\sum_i^{N+L} |\lambda_i(t_k)|)/h \in \mathbb{R}$, which is a measure of the average size of each load change across the network. It provides information about the average magnitude of network loads the attacker must manipulate to trigger the cascade observed.
\end{enumerate}
%  \ale{Explain briefly why do the two metrics have different units.}

\vspace{-0.15 cm}

\subsection{Network emergency responses and cascading failures}
\vspace{-0.1 cm}
\label{sec:emer_res}
Emergency responses (ER) are the systems safety mechanisms that safeguard sensitive network equipment from dangerous deviations in frequency-related profiles. We provide a list of ER employed in the following.  
%and refer the reader to \cite{goodridge2020} and \cite{goodridge2021} for a detailed mathematical description. 
% \hspace{-0.2cm}
\begin{enumerate}[leftmargin=.5cm,labelsep=0.2cm]
    \item \emph{Generation shedding:} we model two independent schemes which disconnect generators from the network: (i) RoCoF-induced generation shedding (RIGS) -- generation is disconnected when nodal RoCoF $|\ddot{\delta_i}|$ exceeds an upper threshold; (ii) over frequency generation shedding (OFGS) -- generation is shed when nodal frequency $\dot{\delta_i}$ exceeds a pre-set upper limit.
    \item \emph{Under-frequency load shedding (UFLS):} this scheme disconnects of 10\% of equilibrium nodal loads when the frequency $\dot{\delta}_i$ falls below a strictly decreasing sequence of four frequency thresholds. 
    \item \emph{Line disconnection:} If the power flowing through an inter-connector line linking different areas in the power grid exceeds a pre-set upper threshold, then the line gets disconnected.
\end{enumerate}

% In the IEEE 39-bus network, we model the disconnection of the lines connecting Areas 1, 2, and 3 when 

%When excess power flow is detected through the inter-connector line, the indicator $\Omega_{ij}$ switches from 1 to 0 for the remainder of the simulation, setting the $ij^{th}$ element of $B_{ij}$ to 0 \cite{goodridge2021}. 

% The ER model inspects the continuous time variables $\dot{\delta}_i(t)$, $\ddot{\delta}_i(t)$ and $\phi_{ij}(t)$ from the power system model~\eqref{eq:network_model} at regular time intervals. 
% Once a criterion for activation is observed, the corresponding ER is activated. This is represented in \eqref{eq:network_model} as a discontinuity, where changes to the relevant input variables (power injection, load or network topology) are applied. Subsequently, the simulation is resumed with the new network parameters. 

%\cite{goodridge2020},
% : lost generation represents foregone revenues for generator operators and load shedding curtails national economic output \cite{Li2017cascades}, while line losses increase inspection and repair costs for transmission operators.

While ERs are intended to arrest large deviations in frequency-related dynamics when coupled with the effects of D-LAAs, multiple disconnection events may transpire on the network  even when such power grids are designed to be $N-1$ secure \cite{goodridge2021}. Such \textit{cascading failures} threaten network integrity and result in significant costs to various network stakeholders. 
We measure the \textit{cascade size} $\mathcal{X}$ resulting from a D-LAA as the cumulative power (in MW) of network components (loads and generators) disconnected during the power grid operation following the LAA. 

%Thus:
% \begin{equation}
%     \mathcal{X}= \sum_i^{N+L}{\big(\chi_i^G\phi_i +0.1\chi_i^L\mathcal{R}_i\big)}
% \end{equation}
% The loss of inter-area transmission lines was not included in this metric as 

	%%%%%%%%%%%%%%%%%%%%%%%%%%%%%%%%%%%%%%
	%		THE NETWORK
	%%%%%%%%%%%%%%%%%%%%%%%%%%%%%%%%%%%%%%

\section{Sampling Methodology for D-LAAs}\label{sc:statmodel}
% In this section, we present the statistical model for the distribution of LAAs and describe the proposed rare-event sampling approach to identify the impactful LAAs. 
% \subsection{Modeling the Unconditional Distribution of LAAs}
% 	At each node $i$ of the Kron-reduced network, we sample a random increase or decrease in the pre-LAA vulnerable load $\chi^{LV}_i$, representing a change power consumed by vulnerable appliances commanded by an attacker. 
	
%A review of the literature revealed no definitive documentation of instances or data-sets of LAAs against power grids. Nevertheless, 

% There are several studies that document the frequency and magnitude of cyber breaches in enterprise networks \cite{Edwards2016}. For instance, \cite{Edwards2016} demonstrates thCumuale size of data breaches can be well captured by the log-normal family of distributions. 
In this work, we apply a sampling methodology to generate various spatiotemporal instances of LAAs (i.e., across the victim nodes and attack intervals). We apply them as inputs to the power system model described in Section~\ref{sec:Prelim} and assess the cascading failures that occur as a result. The detailed models are presented next. 
% We assume an agnostic sampling methodology, intended to explore the effects of various combinations of network and D-LAA characteristics with few assumptions. More specifically: 

\begin{enumerate}[leftmargin=.5cm,labelsep=0.2cm]
    \item We model the \textbf{D-LAA magnitudes} at time $t_0$ at all $L$ vulnerable nodes as independent, uniformly distributed random variables, namely $\bm{\lambda}=(\lambda_1(t_0),\dots,\lambda_L(t_0)) \sim \mathcal{U}[0,\lambda^0_{\max}]^L$, where $\lambda^0_{\max}$ denotes the attack limits. The D-LAA magnitudes at subsequent epochs are then uniquely determined by the dynamics described in~\cref{sec: laa_model}. The wide support for the uniform distribution allows us to capture also extreme scenarios in which an attacker manipulates a large proportion of the load.
    \item We model the \textbf{interval between load changes} using a uniform discrete distribution, i.e. $\mathcal{I}\sim \mathcal{U}[1,T_{\max}],$ where $T_{\max}$ (in seconds) denotes the duration of the dynamic simulations. Note that we do not allow subsequent attacks to be less than $1$ second apart to account for the time an attacker may need to estimate the system frequency to calibrate their next move. 
    \item We sample the \textbf{scenario} $\tau$ uniformly among a set of four, each representative of the intra-day power equilibrium at a different moment of the day (see~\cref{sec:simsettings}).
    \item We model the \textbf{D-LAA-frequency response} as a uniform random variable $C \sim \mathcal{U} [C_{\min}, C_{\max}]$, where $C_{\min}$ and $C_{\max}$ denote practical limits for the D-LAA. 
    % However, all our results were invariant with respect to the variable $C$ in this range, so it has been omitted from our analysis. 
\end{enumerate}
% loall loads at a node   Given the dynamic nature of We model nodal dynamic LAAs Assuming nodal LAAs magnitudes are independent and follow a  distribution, e model $U\in \mathbb{R}^{N+L}  \coloneqq[|u_1|, \dots, |u_{N+L}| ]$ as
% \begin{equation} \label{eq:rho}
%     U \sim  \prod_{i=1}^{N+L} \text{Lognormal} (\mu_i,\sigma_i^2). 
% \end{equation}
We note our sampling procedure is not constrained to the uniform distribution and, in fact, it can be easily extended to accommodate any underlying distribution.
% analysis is not restricted to the log-normal family of distributions, and can be extended to any underlying distribution in a straightforward manner.

%Note that the log-normal distribution models the fact that low-magnitude attacks are more likely to occur, where as LAAs of very large magnitude occur rarely. The underlying distribution $\rho$ is therefore given by:
	 
% 	\begin{equation}
% 	    \rho \sim \prod_{i=1}^{N+L} Lognormal(\mu_i,\sigma_i^2)
% 	\end{equation}
% 	where $\mu$ and $\sigma$ are given in table [xxx].

% 	{\color{red}  Sub: Do we have the subscript $i$ or not?}

%As noted before, the objective of our work is to investigate the attack impact under all potential distributions of LAAs across the victim nodes. 
%To this end, we require a statistical model of LAAs (i.e., the intensity of LAAs at different nodes). 

% \subsection{Unconditional distribution}\label{sec:us}
% Following  the magnitude of LAAs follow a log-normal distribution. Thus $u_i|\nu \sim Lognormal(\mu,\sigma^2)$.% where $\mu =1$ and $\sigma^2 = 4$. 
	
% \subsection{Rare-Event Sampler for LAAs}
% \label{sc:skipping_sampler}
The sampling assumptions for the various D-LAA features outlined in the previous subsection result in an unconditional product density $\rho$ over $\mathbb{R}^{L+3}$, where each vector of attack parameters $\bm{x} = (\bm{\lambda}, \mathcal{I}, \tau, C)$ fully characterizes a D-LAA attack in view of the dynamics we described in~\cref{sec: laa_model}. Since we are interested in studying the cascading failures triggered by D-LAAs, we need to sample attack parameters $\bm{x} = (\bm{\lambda}, \mathcal{I}, \tau, C)$ that results in the activation of at least one network ER. Let $A \subset \mathbb{R}^{L+3}$ be the subset of such ``successful'' attack parameters. We are interested in sampling D-LAA attacks according to the density $\rho$ but \textit{conditionally} on the event $A$. Let $\pi_A$ be such conditional distribution over $\mathbb{R}^{L+3}$, i.e., $\pi_A(x) := \rho(x)\mathbbm{1}_A(x)/ \rho(A)$,
% {\begin{equation*}\label{eq:pi}
%     \pi_A(x) := \frac{\rho(x)\mathbbm{1}_A(x)}{\rho(A)},
% \end{equation*}}%
where $\rho(A)$ is the probability of the event $A$ occurring, and  $\mathbbm{1}_A(x) =1$ if $x \in A$ and $0$ otherwise. Due to the pre-existing network security mechanisms, however, it is very unlikely that a D-LAA would trigger a network ER, which means that $A$ is a rare event with $\rho(A) \ll 1$. Any standard MCMC sampling method would then struggle enormously at sampling from the conditional density $\pi_A$.
%
% {\color{red}this table can be removed and cited from a growing number of references}
{\small \begin{algorithm}[!ht]
\SetAlgoLined
\SetKwInOut{Input}{Input}\SetKwInOut{Output}{Output}
\textbf{Input:} initial state $U_1$;\\
\For{$i = 1,\dots,n$}{
\BlankLine
\DontPrintSemicolon
% 		$\underbar{\ensuremath{u}}\coloneqq \underbar{\ensuremath{u}}_n$;\\
Generate an initial proposal $\ensuremath{Z_1}$ distributed according to the density $q(\ensuremath{y} - \ensuremath{U_i})d\ensuremath{y}$;\\
Calculate the direction $\ensuremath{\Phi}=\left( \ensuremath{Z_1}-\ensuremath{U_i}\right)/\left\Vert \ensuremath{Z_1}-\ensuremath{U_i}\right\Vert $;\\
Generate a halting index $K \sim K_{\varphi}$;\\
Set $k=1$;\\
\While{$\ensuremath{Z}_k \notin C \textbf{\textup{ and }}$ $k< K$}{
    Generate a distance increment $R$ distributed according to $q_{r|\ensuremath{\Phi}}\left(r|\ensuremath{\Phi}\right)$;\\
    Set $\ensuremath{Z}_{k+1}=\ensuremath{Z}_k+ \ensuremath{\Phi} R$;\\
    $k=k+1$;\\
}
Set $\ensuremath{Z}\coloneqq \ensuremath{Z}_k$;\\
Evaluate the acceptance probability:
\begin{equation}\label{eq:ap2}
    \alpha(\ensuremath{U_i},\ensuremath{Z})=
    \begin{cases}
        \min\left(1, \frac{\pi(\ensuremath{Z})}{\pi(\ensuremath{U_i})} \right) & \text{ if } \pi(\ensuremath{U_i}) \neq 0, \\
        1, & \text{ otherwise.}
    \end{cases}
\end{equation}\\
Generate a uniform random variable $V$ on $(0,1)$;\\
\uIf{$V \leq \alpha(\ensuremath{U_i},\ensuremath{Z})$}{$\ensuremath{U}_{i+1}=\ensuremath{Z}$;}
\Else{$\ensuremath{U}_{i+1}=\ensuremath{U_i}$;}
\KwRet{$\ensuremath{U}_{i+1}$}.\;
}
\textbf{Output}: final sample $[U_1, U_2, U_3, \dots]$
\caption{Skipping sampler algorithm}
\label{alg:skip}
\end{algorithm}}

We thus employ the \textit{skipping sampler} MCMC algorithm proposed in~\cite{Moriarty2019} to efficiently draw samples from $\pi_A$. The steps followed are presented in Algorithm~1. As a Metropolis-class algorithm, the skipping sampler can be understood as a two-step procedure consisting of a \textit{proposal step} (Steps 3-12 of Algorithm~1) and an \textit{acceptance/rejection step} (Steps 13-21 of Algorithm~1). We omit a detailed description of Algorithm~1 and provide an intuitive explanation of the working of the skipping sampler next. 
%
%, which determines whether the proposed state $Z$ is included in the final sample, according to a specified acceptance probability. This ensures the distribution of the sample follows the desired target distribution $\pi$. If it is accepted, the proposal is included in the final sample and becomes the starting state for the next proposal step. This procedure is repeated a desired number of times, after which the final sample is returned. 
%

% \vspace{-0.25cm}
\begin{figure}[h!]
    \centering
    \includegraphics[width=0.4 \textwidth]{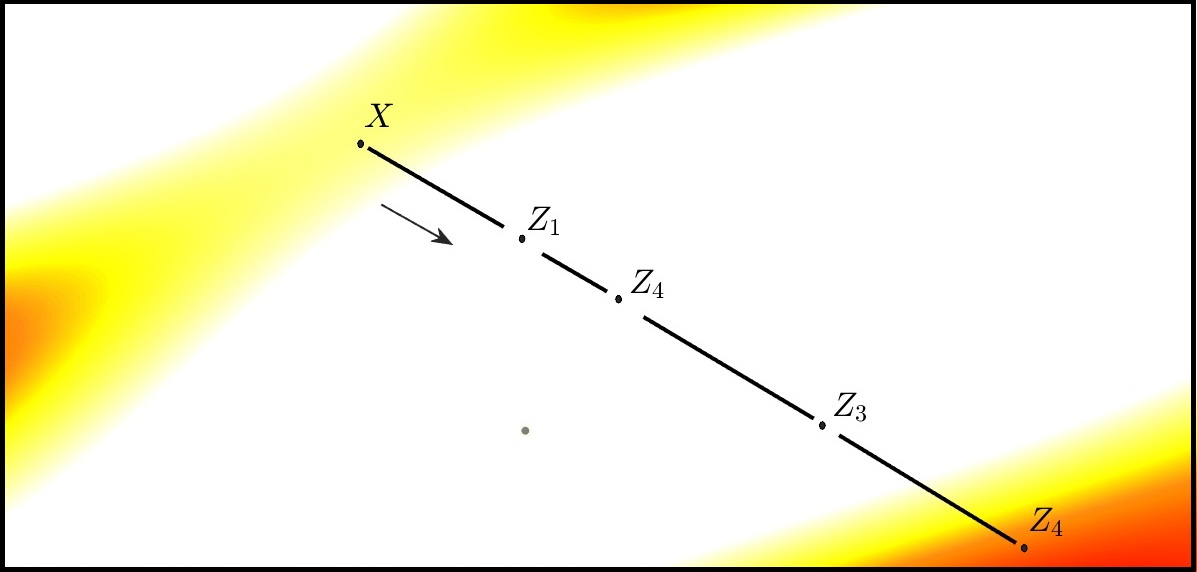}
    % \vspace{-0.21cm}
    \caption{{\small  Illustration of a skipping proposal in $\mathbb{R}^2$ attempting to sample from the subset $A$ (coloured regions). Starting at $X$, the initial random-walk proposal $Z_1$ would be rejected since $Z_1 \not\in A$ by a classical MCMC method. Instead, the skipping proposal generates random distant increments and finds new $Z_k$'s in a linear fashion (`skipping') along the original direction until $Z_k \in A$ or updates are halted.}}
    \label{fig:skipping}
\end{figure}
% \vspace{-0.2cm}
In essence, the skipping sampler improves the conditional sampling from the subset $A$ by using a specialized proposal step that `skips' over proposals in $A^c$ (the set of D-LAAs parameters that do not lead to the activation of an ER) until a point $\bm{x} \in A$ is sampled, or the skipping process is halted in a randomized fashion (see Fig.~\ref{fig:skipping} for an illustration). 
A key advantage of the skipping sampler is its ability to efficiently transition between potentially disconnected components of the subset $A$ and different modes of the conditional density $\pi_A$. This is particularly important given the nonlinear dynamics of the network and the presence of heterogeneous ER mechanisms, which can result in a disconnected subset $A$ of successful attacks. With a standard MCMC sampling strategy, exploring multiple disconnected components of $A$ would be challenging, leading to incomplete exploration of the range of attack parameters and results heavily dependent on the starting configuration. In contrast, the skipping sampler is designed to effortlessly navigate across different connected components of $A$. In the context of D-LAA attacks, this means the skipping sampler can generate samples with diverse attack parameters and features, allowing for the discovery of all major network vulnerabilities.

%Denoting the current state $U_n$, if the initial proposal $Z_1 \notin D$, we update the initial proposal (or `skip') by adding an independent random distance increment $R_2$ in the direction $\Phi = \frac{Z_1 - U_n}{|Z_1 -U_n|}$, where $R_k$ has the conditional distribution of $\left\Vert Z_1-U_n\right\Vert$ conditioned on $\Phi$. 

%-------------------------------------------------------------------
%
%	CASE STUDY SETUP
%
%-------------------------------------------------------------------
\vspace{-0.1cm}
\section{Simulations}\label{sec:Sims}

\subsection{Simulation Settings}
\label{sec:simsettings}

% \vspace{-0.15cm}

We implement the power system dynamics and D-LAAs on a Kron-reduced version of IEEE 39-bus test network, consisting of $N = 10$ generation buses (2 of which also have loads present) and $L = 17$ pure load buses \cite{ieee39reference}.

%%% IEEE-39 bus figure 
% \begin{figure}[!ht]
%     % \vspace*{-0.75cm}
% % 	\hspace*{-0.5cm}
%  	\centering
% % 	 \qquad
% % 	 \subfloat[][Average cascade size by $\nu$]
% 	 {\includegraphics[width=0.38\textwidth]{conference paper charts/ieee 39 network.jpg}    \label{fig:ieee39 network}
%  }
% 	 \caption{Schematic of the IEEE39 network showing 3 areas operated by different utility operators}
%  \end{figure}
%%%%%%

At $t = 0^-$, the system equilibrium power balance is modelled to be in one of four load-scenario states $\tau \in \{1,2,3,4\}$, corresponding to four peaks of the diurnal load cycle of a typical European network, colloquially labelled \emph{night, morning, afternoon \text{and} evening} respectively. Equilibrium active loads (and power) are calculated as proportional changes to the published equilibrium power balance of the IEEE 39-bus network \cite{ieee39reference}, whose relative proportions of $\{0.4,1,0.85,1.3\}$ for $\tau=1,\dots,4$ respectively are based on the UK power grid's cycle \cite{ukdiurnalcycle2018}. Broadly speaking, evening and morning periods have the highest load, whereas night-time has the lowest load conditions. 
The choices of protection system parameters are such that the network is $N-1$ secure, such that the loss of a single component (generator, load, or line) in the absence of any other disturbance does not trigger a subsequent ER. 
% The four initial conditions associated with the 4 equilibrium states of the above system of equations, denoted $\delta_i(0)$, $E_i (0)$, $\rho_i(0)$,$P_i^L$ and $P^G_i$ are determined numerically such that $\delta_i \approx 0$ for each equilibrium state. 

To generate dynamic LAAs associated with the disconnection of network components, we utilise the \textit{skipping sampler algorithm} as follows. Starting from an initial nodal LAA load change $\bm{\lambda}\in \mathbb{R}^L$, we sample 
During each proposal step, we sample a tuple $\bm{x} = (\bm{\lambda}, \mathcal{I}, \tau, C)$ consisting of: the initial nodal LAA vector $\bm{\lambda}$, an interval between attacks $\mathcal{I}$, a scenario $\tau$ and a frequency response $C$. The sampling parameters are set to $\lambda^0_{\max} = 1000~\text{MW}, T_{\max} = 60$~s, $C_{\min} = 0.5$  and $C_{\max} = 5$. We then apply the components of $\bm{x}$ as an input to the power system model \cite{goodridge2021} at time $t= 0$, with frequency dynamics simulated for $60s$ using MATLAB. We conduct $n=100,000$ proposals, which resulted in a final sample of $S \approx 16,500$ dynamic LAAs that activated at least one ER. This corresponds to an acceptance rate of 16.5\%, which is within the optimal 15--48\% acceptance rate range~\cite{Gamerman2006}. From each sample, we estimate the metrics of the cascade size $\mathcal{X}$ in MW and the average magnitude of loads changed $\{\Sigma_i(\lambda)\}_{i=1}^L$.
\vspace{-0.1 cm}

\subsection{Simulation Results}
% \vspace{-0.1 cm}
% \subsection{Consequence of dynamic LAAs on the IEEE 39 network}

% {\color{blue} 1. There's legend in some plots, and some others dont \\ 
% 2. Remove titles on top of each plot, its redundant}

% Evaluating cascade sizes resulting from simulated dynamic load-altering attacks (D-LAAs) applied to the IEEE 39 network,  highlights the particular susceptibilities of the network to malicious attacks against network loads.

\subsubsection{{\textbf{Overview of Cascade Sizes}}}

\begin{figure}[!h]
    \vspace*{-0.2cm}
    \hspace*{-0.7cm}
    \centering
    {\includegraphics[width=0.4\textwidth]{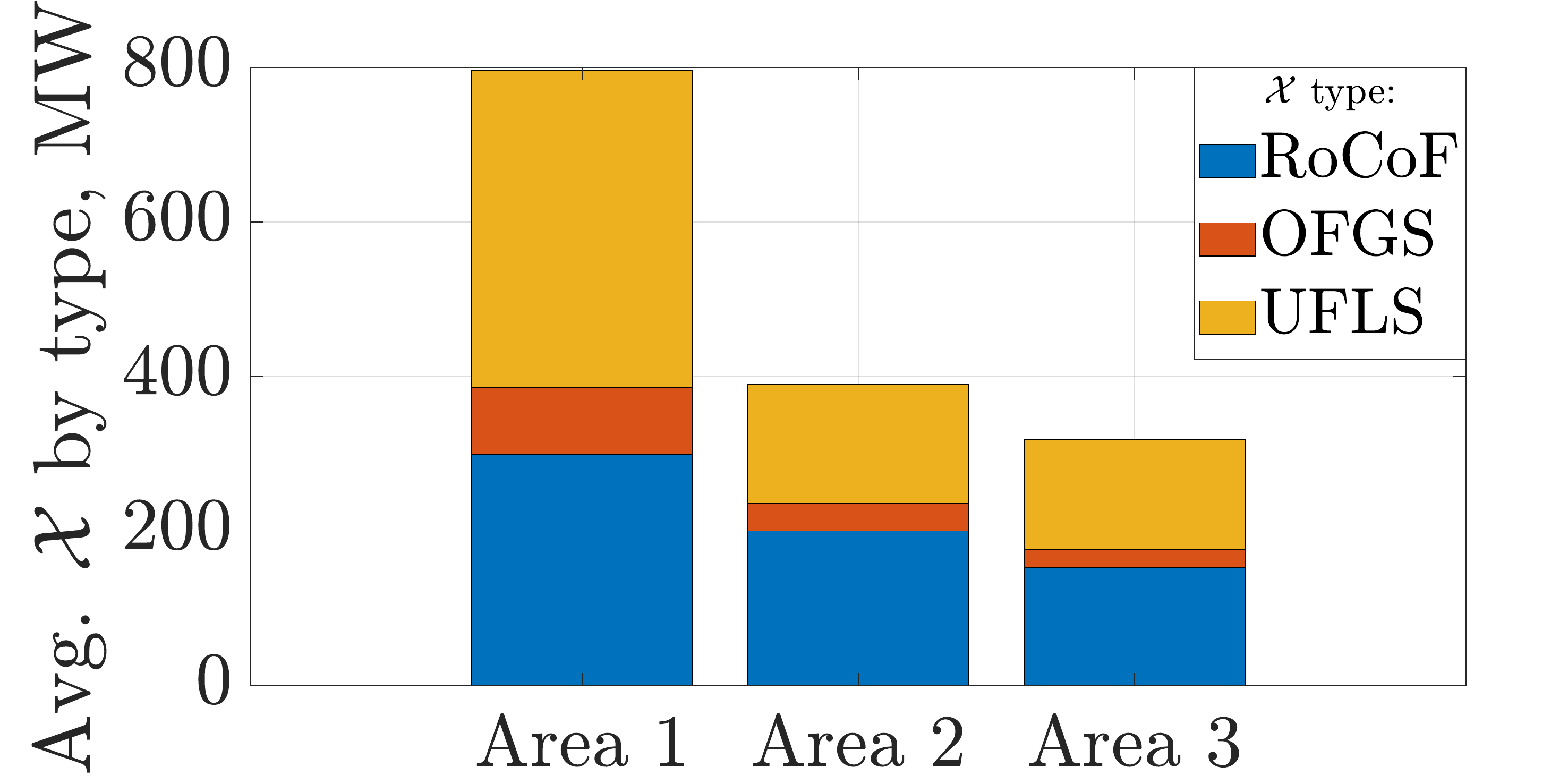}\label{fg:ieee areas}}
     \vspace{-0.2cm}
    % \centering
    \caption{{\small Average cascade size by area for the IEEE 39-bus network.}} 
    \label{fig:network analysis}
    % 	\vspace*{-0.35cm}
\end{figure}

{The results shown in Fig.~\ref{fig:network analysis} indicate that cascades in the system are on average primarily caused by load shedding and RoCoF-induced generation disconnections. These findings have three important implications: (1) globally, the IEEE-39 bus system is most vulnerable to D-LAAs which increase loads, resulting in the preponderance of UFLS events. (2) The network exhibits greater resilience to LAA-induced net reductions in loads, as evidenced by the relatively small proportion of cascades associated with OFGS-related generation disconnections. (3) Among all types of disconnections, individual UFLS events are the most prevalent in the sample. Although each UFLS event only sheds a small magnitude of load in MW, they can be activated multiple times at each node, thus the ~700MW of load disconnections in fact consists of a large number of individual UFLS events. In the following sections, we explore how both network and LAA parameters contribute to these results.}
\vspace{-0.2cm}

% \subsubsection{Impact of Attack \textcolor{blue}{Characteristics}}
% {\color{blue} We investigate how the LAA characteristics impact cascade sizes induced by D-LAAs.}

% We explore the relationship between $\mathcal{X}$ and the vulnerability ratio $\nu$ and the interval between LAA updates $\mathcal{I}$.

 % The impact of any load-altering attack was expected to depend intrinsically on the  proportion of network loads vulnerable to cyber-based manipulation.
 
\begin{figure}[!h]
    \vspace*{-0.1cm}
 	\hspace*{-0.5cm}
 	\centering
% 	 \qquad
% 	 \subfloat[][Average cascade size by $\nu$]
	\includegraphics[width=0.45\textwidth]{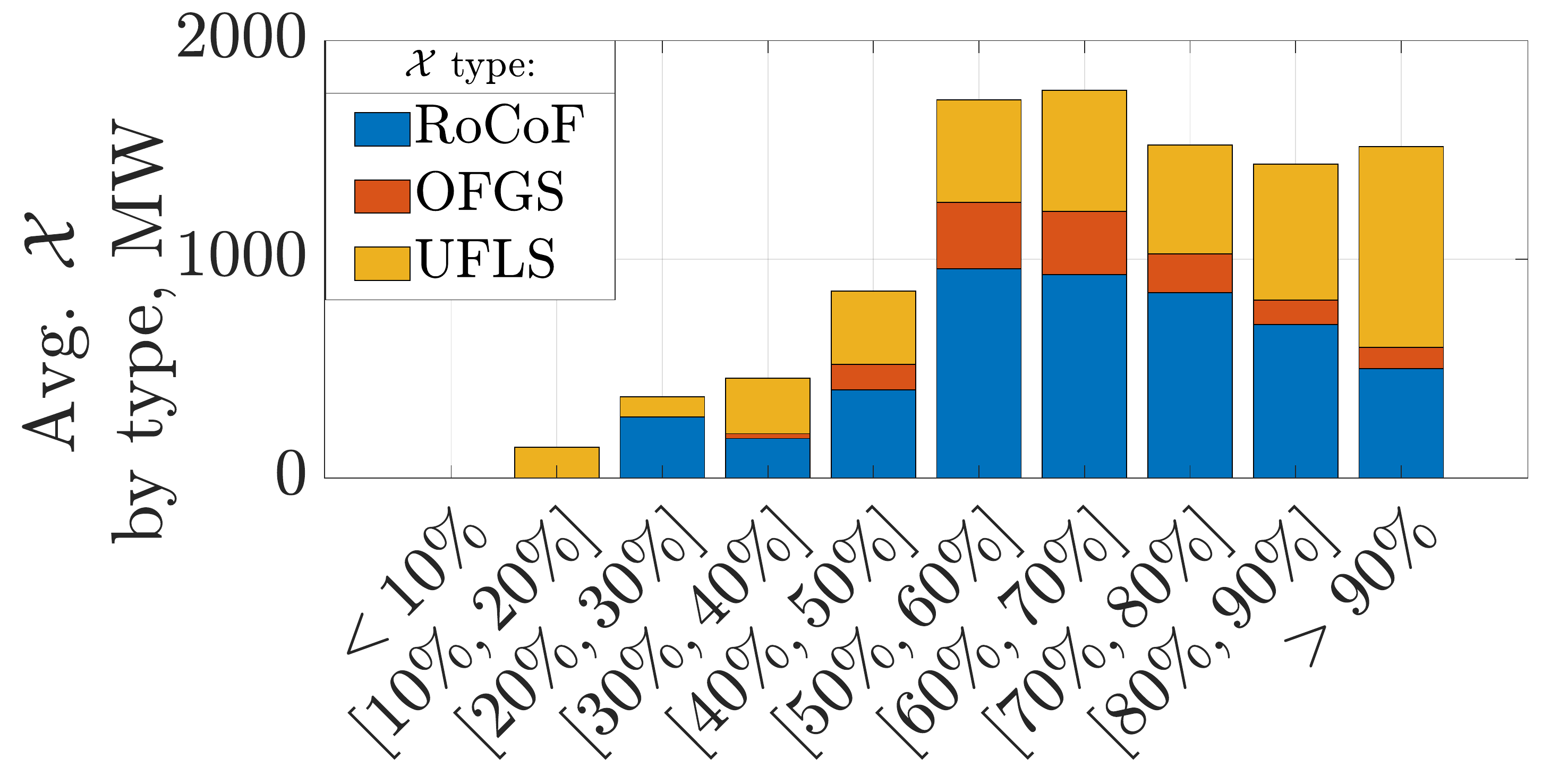}
	 \vspace{-0.2cm}
	 \caption{{\small Average cascade size (in MW) by $\nu.$}}
  \label{fg:cascade_nu}
  \vspace{-0.2cm}
 \end{figure}

\textbf{Impact of the vulnerability ratio ($\nu$) on cascade size:} The vulnerability ratio reflects the proportion of the total load %($12.9$ GW for the IEEE-39 bus system) 
an attacker can potentially control at each node. Hypothesising that $\nu< 30\%$ may be possible for current and near-future power systems, we observe that disconnections in this regime consist of UFLS and RIGS. As per Fig.~\ref{fg:cascade_nu}, the IEEE-39 network demonstrated resilience against D-LAAs for $\nu\le 10\%$, with no components disconnected in this regime as the attacker lacks sufficient leverage to disrupt network operations. For $\nu\in[10,20)\%$, disconnections are exclusively UFLS, reinforcing the network's susceptibility to this form of disconnection following a D-LAA. Further, the network exhibits greater resilience against generation shedding, with RIGS requiring $\nu\ge 20\%$, and OFGS disconnections $\nu\ge30\%$.

%%% I propose that we defer defense-based discussion to future work %%%%
%% {\color{red}MG: These results have implications for network planners as the proportion of vulnerable IoT devices slowly increases. For low penetration rates (e.g $\nu<20$), improvement in UFLS protection systems may improve network resilience to malicious cyber threats via IoT loads. For larger rates of penetration, the focus of planners must turn to securing network generators. The analysis of mitigation techniques has been left for a future, dedicated article. }
%%%%%%%%%%%%%%%%%%%%%%%%%%%%%%%%%%%%%%%

% mg- removed this section which analyses general trends in cascades and vulnerability outside of relevant realistic ranges. May reinclude it if space opens up.

A trend analysis reveals when $\nu \in [10\%, 60\%]$, there exists a positive relationship between cascade size $\mathcal{X}$ and $\nu$ -- as the attacker gains more authority over loads nodes of the network, each effective load change in the dynamic LAA sequence can be larger in magnitude, inducing larger frequency deviations and thus larger cascades. In this vulnerability range, the increase in cascade size is driven primarily by the disconnection of large generating units, mainly due to RoCoF violations, with minor contributions from over-frequency generation shedding. 

For $\nu\ge60\%$, we observe stagnation in the growth in total cascade sizes. Simultaneously, the average load shed increases. This behaviour demarcates a phase transition in the susceptibility of the network- for $\nu<60\%$, cascades are characterised by the growth of generation disconnections, while for $\nu\ge 60\%$, average generation disconnections decrease, while load disconnections increase to eventually dominate disconnections as the attacker gains greater leverage over network loads.

%% Removed Avg. cascade size vs interval %%%%%%%%%% time plot
% \begin{figure}[h!]
% \vspace{-0.3cm}
% \centering
%       {\includegraphics[width= 0.48\textwidth]{conference paper charts/cascade mw by I.pdf}}
%       \caption{Average cascade size (in MW) by $\mathcal{I}$}
%       \label{fg:cascade I}
% \end{figure}
% \vspace{-0.1cm}
% Figure~\ref{fg:cascade I} reveals the distribution of average cascade sizes with respect to $\mathcal{I}$ is bimodal, with the average size of cascades $\mathcal{X}$ highest when the IEEE 39 network is exposed to D-LAAs with very short intervals ($\mathcal{I}\le 5s$) or very large intervals ($\mathcal{I}\ge45s$) between load changes. Both modes are dominated by the disconnection of large generators, primarily driven by RoCoF violations. This counter-intuitive result is best understood by investigating the influence of both $\mathcal{I}$ and $\mu_{\lambda}$ on the cascade sizes $\mathcal{X}$ reported in the IEEE39 network, visualised in figure~\ref{fig:cascadesize_mu_I}. 
%%%%%%%%%%%%%%%%%%%%%%%%%%%%%%%%%%

%In contrast to S-LAA strategies, where an adversary can only manipulate two characteristics of each attack, namely, the magnitude and location in the network, adversaries employing dynamic attack methodologies can also exploit a third temporal dimension by varying the interval between individual load changes, $\mathcal{I}$. 

\begin{figure}[!h]
\vspace*{-0.2cm}
\hspace{-0.1cm}
\centering
\centerline{\includegraphics[width=0.45\textwidth]{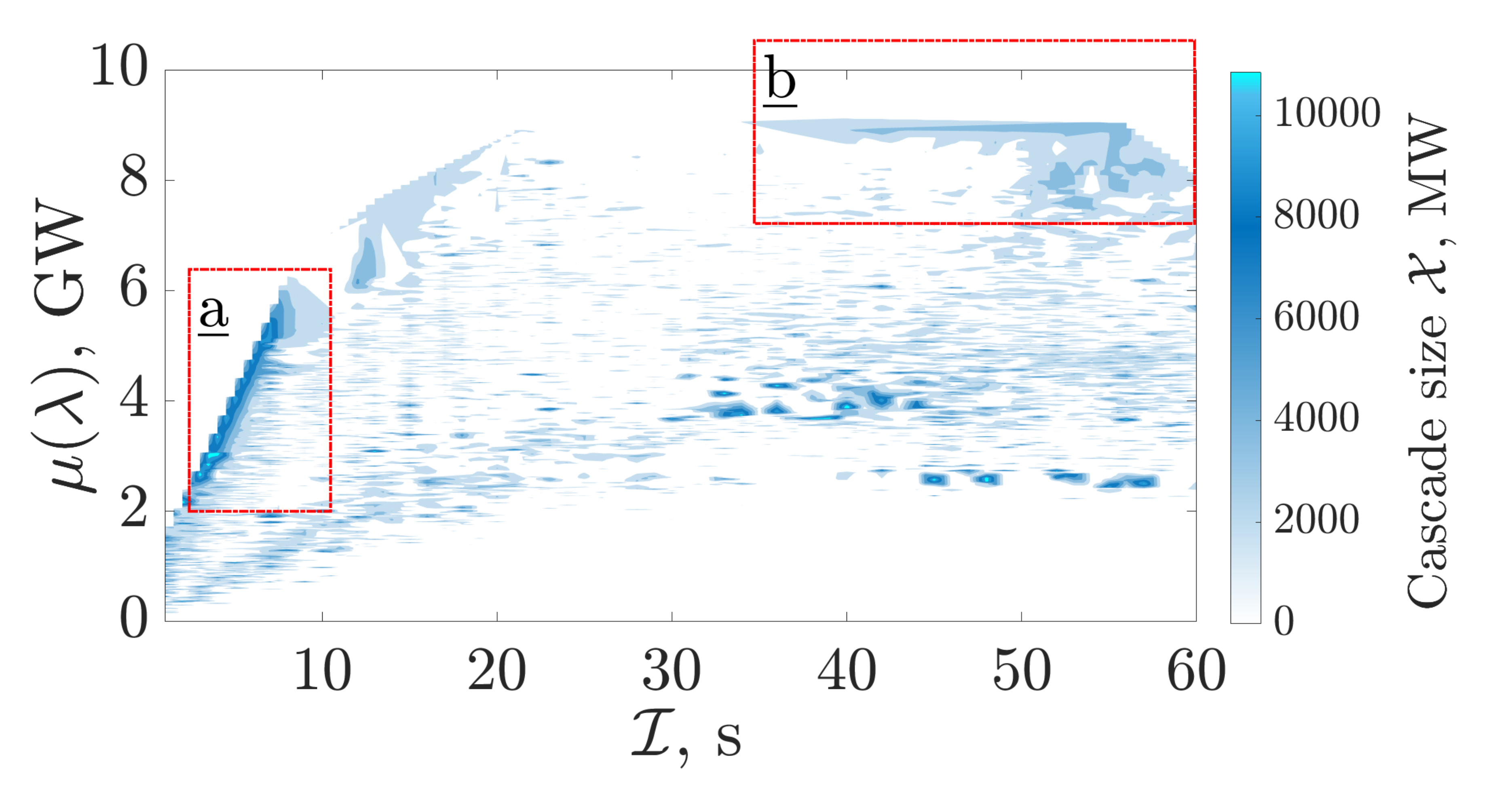}}
\vspace{-0.8cm}
\caption{{\small Color map showing the cascade size $\mathcal{X}$ (in MW), with respect to the interval between dynamic LAAs $\mathcal{I}$ (s) and the average size of each LAA's load change $\mu_{\lambda}$, (GW).}}
\label{fig:cascadesize_mu_I}
\end{figure}

{\textbf{Interval between dynamic load changes} $\mathcal{I}$:} 
Fig.~\ref{fig:cascadesize_mu_I} reveals two regimes particularly susceptible to large cascades: (\underbar{a}) rapidly changing ($\mathcal{I}<10s)$, smaller-magnitude ($\mu(\lambda)<4 GW$) LAAs and (\underbar{b}) static ($\mathcal{I}>50s$), large magnitude ($\mu(\lambda)>7 GW$) LAAs. D-LAAs in region \underbar{a} result in larger cascades with an average size of 9,000MW and are more threatening to network integrity.
Conversely, the region \underbar{b} is associated with large, static attacks, resulting in an average cascade size of ~4,000MW. Thus, by deploying multiple attacks, informed either by quasi-real-time or simulated frequency data, an attacker can exacerbate frequency and RoCoF deviations in an intentional and strategic manner, inducing cascades using smaller magnitude D-LAAs when compared to S-LAAs. 
 % \vspace{-0.275cm}
The efficacy of this strategy is observed in Fig.~\ref{fig:cascadesize_mu_I}, where, for example, when the interval between attacks is $50s$, the attacker must be able to manipulate at least ~1,550MW of network loads to trigger a disconnection event. In comparison, when $\mathcal{I}=10s$, the attacker needs only manipulate $\mu(\lambda)\ge300$MW of loads across the network in a coordinated fashion to trigger a disconnection event. For reference, 300MW of loads may represent 170,000 typical space heaters or 45,000 charging EVs, a magnitude of loads possible in modern networks.
In general, the lower boundary of Fig.~\ref{fig:cascadesize_mu_I} represents the \textit{critical D-LAA characteristic threshold} -- the minimum average load change which must be manipulated for each attack interval $\mathcal{I}$ to induce a disconnection event. The positive gradient of this boundary establishes that smaller magnitude D-LAAs require shorter attack intervals to trigger similar-sized cascades.

\subsubsection{Impact of Power Grid Operating Conditions}
Last, we consider the impact of power grid operating conditions.

{\bf Inter-Area Power Balance:}
Decomposing results by the areas of the IEEE-39 network, cascade sizes $\mathcal{X}$ are related to the initial ($t < 0^-$) net generation imbalance of each area. Area 1, with excess demand, is particularly susceptible to large cascades,  half of which are attributable to a large number of loss of load events. This is related to the network's susceptibility to D-LAAs which increase loads -- the excess demand profile of Area 1 is exacerbated by such D-LAAs, triggering UFLS events. Area 3, conversely, with near parity between generation and demand, experiences the smallest cascade sizes on average. Being less dependent on other areas to maintain its power balance, it is more resilient to D-LAAs.
% \vspace{-0.45cm}
\begin{figure}[h!]
    \hspace*{-0.7cm}
    \centering
    {\includegraphics[width=0.42\textwidth]{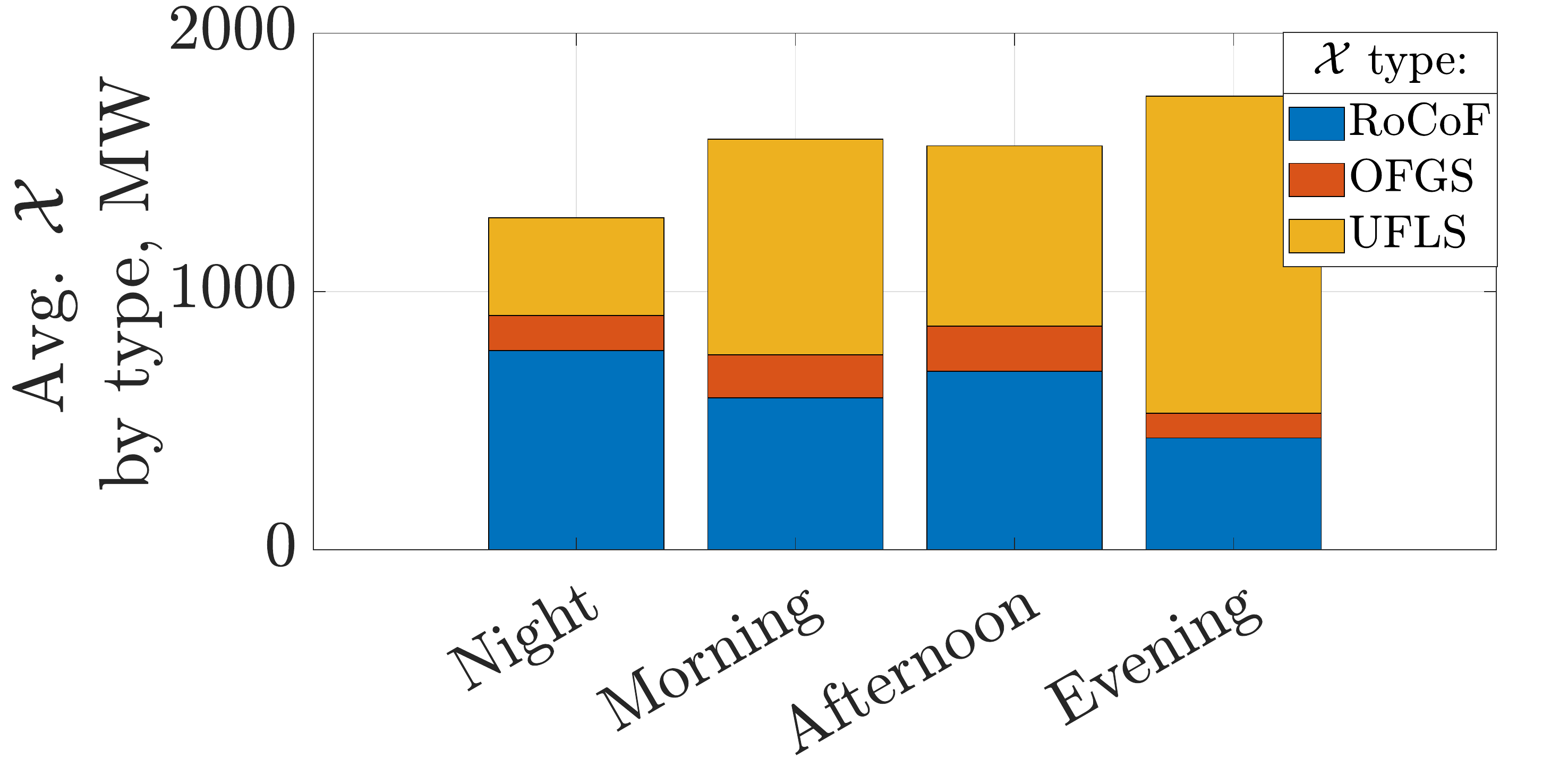}}
    \vspace{-0.29cm}
        \caption{{\small Average cascade size (in MW) by scenario $\tau.$}}
    \label{fg:cascade_tau}
\end{figure}
% \vspace{-0.35cm}

{\bf Network Load Conditions:}
% Next, we investigate the size of cascades $\mathcal{X}$ induced by D-LAAs with the network load conditions. 
%again, preamble is not necessary
On average, $\mathcal{X}$ is minimised during nadir demand (denoted in Fig.~\ref{fg:cascade_tau} as \emph{night}). In this scenario, RIGS dominate cascades, as adversaries induce large changes in frequency through a combination of frequent, alternating changes in loads, or a few, large magnitude increases in load changes, referencing the network's susceptibility to increases in loads. 
Conversely, $\mathcal{X}$ is maximised during periods of peak demand. In this scenario, with most of the network's generating capabilities deployed to serve demand, the network becomes more susceptible to UFLS events from relatively small, positive D-LAA shocks. This is observed in Fig.~\ref{fg:cascade_tau}, where UFLS dominates cascades during the evening, peak demand period. Note that, while load disconnections in MWs will naturally be higher during peak demand, the number of UFLS events, which is invariant to the load scenario, is also maximised during the evening (not shown in the figure). 

%  during evenings, the network becomes more susceptible to UFLS events smaller, positive load changes which increase network loads beyond the network's immediate generation capabilities. \ale{weird sentence, please revise}

% Evaluated by the nodes of the IEEE 39 network, the largest cascades occur at nodes 2, 3 and 10, all of which are in Area 1. Of note, node 10 is a generation node with co-located loads. Any loss of this generator unit, due to RoCoF or OFGS, may lead to local cascading UFLS disconnections, given the pre-existing generation deficit inherent to the area. 

% \subsection{Dynamic LAA Attack Strategies to induce disconnections}

\vspace{-0.1 cm}

\section{Conclusions and Future Research}
\label{sec:Conc}
% \vspace{-0.1 cm}
In this work, we apply a rare-event sampling approach to assess how network variables coupled with LAA parameters influence the size of cascading failures (in MW) in the IEEE39 network. With respect to network variables, our results indicate that the average cascade sizes are larger during peak demand periods and the areas with significant excess demand are particularly vulnerable to cascades. With regard to D-LAA parameters, our results show that the network is resilient against any disconnection event when the proportion of load vulnerable to LAAs is less than  $10\%$ of the base load, while a greater amount
of vulnerable load leads to increasing cascade sizes. Crucial however is the impact of varying the interval between attacks, with results clearly highlighting that shorter intervals between attacks enable an attacker to trigger larger cascades while manipulating a smaller quantity of vulnerable loads, within the range of IoT penetration in the present and near future networks. This exposes a key attack strategy that can be exploited by an adversary with access to network data. Our future work includes exploring optimum strategies and methodologies to mitigate the impact of network D-LAAs, including the usage of battery energy storage systems, the tuning of protection systems, and line parameters optimization. 
\bibliographystyle{IEEEtran}
\bibliography{IEEEabrv,bibliography}

\end{document}